\documentclass[aps,prl,twocolumn,groupedaddress]{revtex4}

\usepackage{graphicx}
\usepackage{dcolumn}
\usepackage{bm}
\usepackage{epstopdf} 

\newcommand{\ket}[1]{\left|#1\right\rangle}
\newcommand{\bra}[1]{\left\langle#1\right|}

\begin{document}

\title{Coherence times of dressed states of a superconducting qubit under extreme driving}

\author{C.M.~Wilson, T.~Duty, F.~Persson, M.~Sandberg, G.~Johansson, P.~Delsing}

\affiliation{Microtechnology and Nanoscience, MC2, Chalmers University of Technology, SE-412 96
G\"oteborg, Sweden}

\date{\today}

\begin{abstract}
In this work, we measure longitudinal dressed states of a superconducting qubit, the single Cooper-pair box, and an intense microwave field.  The dressed states represent the hybridization of the qubit and photon degrees of freedom, and appear as avoided level crossings in the combined energy diagram.  By embedding the circuit in an rf oscillator, we directly probe the dressed states.  We measure their dressed gap as a function of photon number and microwave amplitude, finding good agreement with theory.  In addition, we extract  the relaxation and dephasing rates of these states.
\end{abstract}


\maketitle


When matter and light interact at the quantum level, in the form of atoms and photons, it is often no longer possible to clearly distinguish the individual contribution of each to the overall behavior of the system.  This mixing of the aspects of light and matter can then be described in terms of dressed states of the atoms and photons \cite{CohenTann}.  These dressed states have become an essential concept in many fields of physics.  Recently, they have also been invoked to explain the behavior of electrical circuits operating in the quantum regime  \cite{DelftQED, YaleCavity}.  In this context, known as circuit quantum electrodynamics (QED), we directly measure a class of states, longitudinal dressed states (LDS), that have received little experimental attention in the past.  We create these states by illuminating an artificial atom made from a nanofabricated superconducting circuit with microwave photons.  We then observe the interaction of the dressed states and a radio-frequency (rf) oscillator.  This measurement scheme allows us to directly map the dressed energy diagram and extract the relaxation and dephasing times of the states.  LDS are the natural description of a strongly driven superconducting quantum bit (qubit), and may have applications in the field of quantum information processing.  

Significant advances have been made in the development of engineered systems that exhibit coherent quantum properties.  In the new field of circuit QED, these artificial atoms have recently been used to not only reproduce results of atomic physics and quantum optics, but to explore regimes previously inaccessible to traditional experiments \cite{YaleNumber}.  Here, we use circuit QED techniques to directly study LDS over a wide range of drive strengths, including the extreme driving regime where the driving field is much stronger than the polarizing field.  In atomic systems, the field strengths required for this are technically difficult to achieve and often exceed the ionization threshold of the atoms.  The field geometry studied is also unusual. In a typical atomic experiment, a strong, static field is used to polarize the atomic spins under study.  A relatively weak ac field, aligned perpendicular to the polarizing field, is then used to drive the spins.  This transverse field geometry in fact implies that the atomic and photon spins are aligned, leading to a variety of selection rules based on the conservation of angular momentum.  In our experiment, the driving field is aligned parallel to the polarizing field, relaxing the selection rules, leading to a qualitatively different energy diagram.  The flexibility of nanofabrication also makes it straightforward to simultaneously couple multiple fields to our artificial atom.  We exploit this fact to measure the LDS, directly coupling them to an rf oscillator.  The inherent tunability of the dressed states allows us to explore both the resonant and dispersive coupling regimes.  This enables us to map the dressed energy diagram and extract the relaxation and dephasing times of the states.  

The direct observation of these LDS is of fundamental interest, but also has potential applications in quantum information processing.  For instance, several schemes have been proposed to implement tunable coupling of superconducting qubits through their dressed states \cite{Rigetti, NoriDS}.  These dressed state schemes are attractive because they allow the qubits to remain biased at symmetry points, where they have their longest coherence times \cite{Quantronium}.  

Earlier work with longitudinal driving has shown multiphoton Rabi oscillations between the undressed states of Rydberg atoms\cite{MultiRydberg} and a superconducting circuit\cite{NECMultiRabi}.  A number of recent experiments have also examined superconducting circuits under extreme driving \cite{JenaLZ, MITLZ, FinnishLZ}.  These experiments were interpreted in terms of interference between multiple Landau-Zener transitions in the qubits.   Ref. \cite{MITLZ} could also be interpreted as multiphoton Rabi oscillations, although only the time averaged population of the undressed states was measured.  For the parameter values explored in \cite{FinnishLZ}, the resonances due to different photon numbers overlap, making an interpretation in terms of dressed states less natural.  

\begin{figure*}[t]
\includegraphics[width = 2.1\columnwidth]{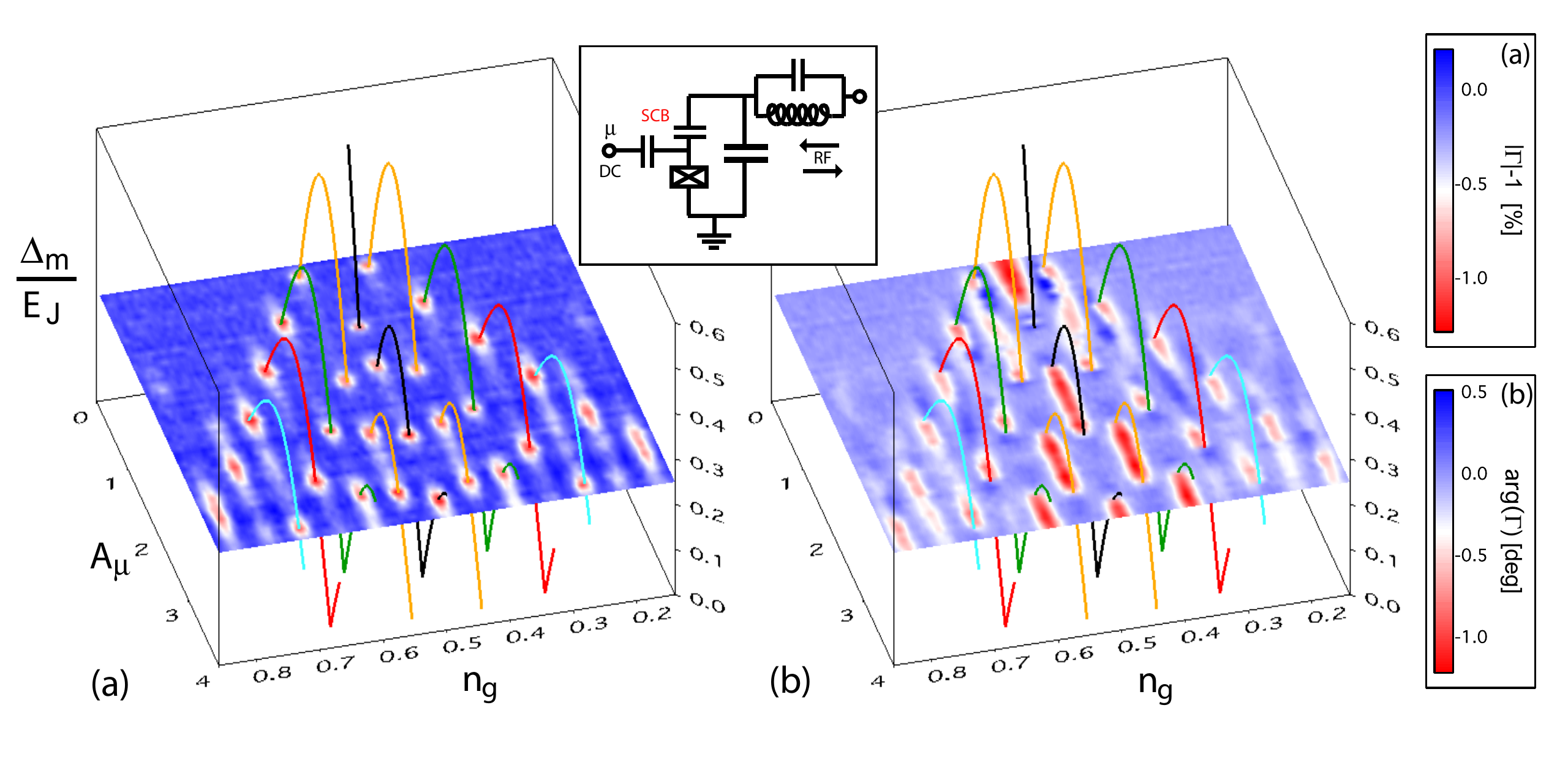}
\caption{RF response of the strongly driven SCB coupled to an oscillator.  The oscillator is formed from an inductor, with stray capacitance, in series with the pad capacitance of the sample (see inset).  It has a resonance frequency, $f_c = 647$ MHz, and a quality factor $Q \approx 40$.  The color images represent the rf reflection coefficient of the coupled system, $\Gamma$, as a function of dc gate charge, $n_g$, and microwave amplitude, $A_{\mu}$.  (Higher $A_{\mu}$ is in the foreground.) (a) is the magnitude response, $\left|\Gamma\right|-1$, and (b) is the phase, $\varphi = \arg(\Gamma)$.  The response is plotted as a plane in a cube along with traces that show the predicted value of the dressed gap $\Delta_m(A_{\mu})$, normalized to $E_J$, for different photon numbers up to m=4.  The z-value at which the response plane is plotted corresponds to the normalized rf probe frequency $hf_c/E_J$.   The data are clearly explained by formally treating the dressed states as tunable two-level systems.  In (a), we clearly see that there is resonant absorption of the rf probe (red spots) at each point that $\Delta_m(A_{\mu }) = hf_c$.  The phase shows the dispersive shift of the resonator frequency when the dressed states are off-resonant.  There are a total of only 2 free parameters, $E_J$ and $\gamma_{\mu}$, for all traces in both panels.  The microwave frequency was 7 GHz and we extract $E_J/h = 2.6\textrm{ GHz}$.}
\label{MagPhase}
 \end{figure*}

Our artificial atom, the single Cooper-pair box (SCB), is composed of a superconducting Al island connected to a superconducting reservoir by a small area Josephson junction \cite{ButtikerQC}.  The excited (ground) state of the SCB represents the presence (absence) of a single extra Cooper-pair on the island.  The dynamics of the SCB can be described by a formal analogy to a spin-1/2 particle \cite{SaclaySCB}.  The  Hamiltonian of the SCB coupled to the driving microwave field is $H = -\frac{1}{2}E_{Ch}\sigma_z - \frac{1}{2}E_J\sigma_x+ \hbar\omega_{\mu} a^{\dagger}a + g\sigma_z(a + a^\dagger)$, where $\sigma_i$ are the Pauli spin matrices and $a^\dagger$, $a$ are the creation and annihilation operators for the microwave field.  The first two terms are the uncoupled SCB Hamiltonian, where $E_{Ch} = E_Q(1-2n_g)$ is the electrostatic energy difference between the ground and excited state of the qubit and $E_J$ is the Josephson coupling energy.  Here $E_Q = (2e)^2/2C_{\Sigma}$ is the Cooper-pair charging energy, $C_{\Sigma}$ is the total capacitance of the island, and $n_g = C_g V_g/2e$ is the dc gate charge used to tune the SCB.  For this experiment, we in fact use a two junction SQUID to connect to the island, which allows us to tune $E_J$ with a small perpendicular magnetic field.  The third term represents the free driving field, and the last represents the coupling,  with strength $g$, between the SCB and the microwave amplitude, $n_{\mu}  \equiv \gamma_{\mu} A_{\mu}/2e$.  Here we define $A_{\mu}$ as the microwave amplitude at the generator and $\gamma_{\mu}$ as the total effective microwave coupling.  We define $A_{\mu}$ such that the generator power in dBm is $20\log(A_{\mu})$.  In the spin-1/2 analogy, $E_{Ch}$ represents the static polarizing field aligned along the z-axis and $E_J$ is a static perturbing field along the x-axis.  The driving field is aligned along the z-axis, giving a longitudinal geometry.

\begin{figure}
\includegraphics[width=1\columnwidth]{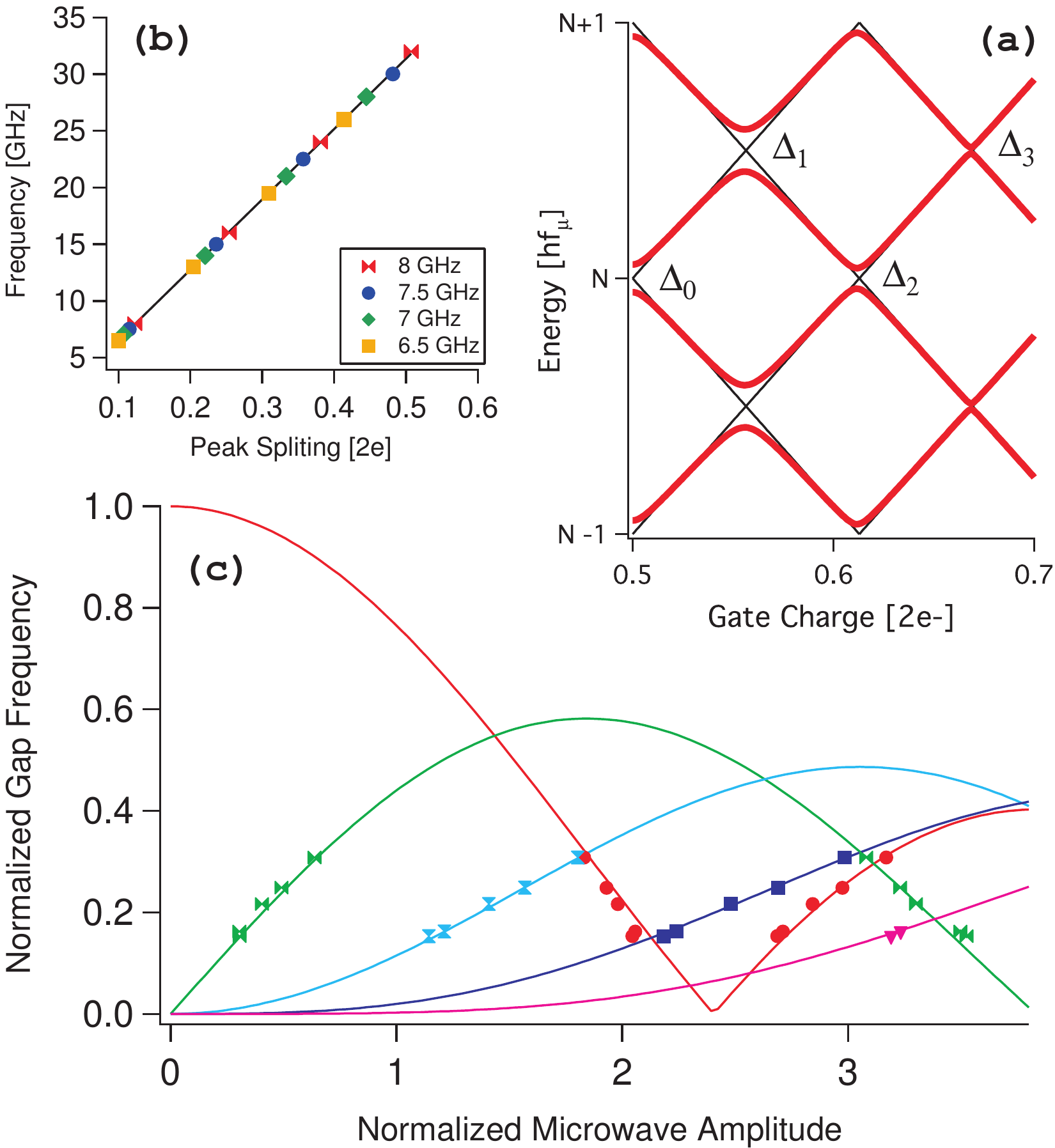}
\caption{(Color Online) Dressed state energy diagram inferred from the magnitude response.  (a) Energy diagram of the combined SCB-photon system.  The thin black lines are the energy bands for $E_J = 0$, showing the degeneracies of states differing by $m=0,1,2,\ldots$ photons regularly spaced in gate charge.  The thick red lines show the degeneracies lifted for finite $E_J$.  The dressed gaps, $\Delta_m$ are functions of the normalized microwave amplitude, $\alpha$.  (b) Position in $n_g$.  The assumed transition frequency, $m\hbar\omega_{\mu}$, versus the splitting in $n_g$ of symmetric resonances.  The data is taken from  Fig. \ref{MagPhase}a and similar data for different $\omega_{\mu}$.  We see that all the points fall on one line with a slope of $E_Q = 62 \textrm{ GHz}$.  (c)  The dependence of the normalized $\Delta_m/E_J$ on  $\alpha$.  We plot data for different values of $E_J/h$, ranging from 2.1-4.2 GHz. The data points are the positions of the resonances in $\alpha$,  plotted at the normalized probe frequencies, $hf_c/E_J$.}
\label{ManyEj}
\end{figure}

With $E_J = 0$, charge states are not mixed and $H$ can be diagonalized exactly  \cite{NECMultiRabi, CohenTann}.  The eigenstates form our dressed state basis 
\begin{equation}
\ket{\pm;N}=\exp[\mp g (a^\dagger-a)/\hbar\omega_{\mu}]\ket{\pm}\ket{N}
\label{states}
\end{equation}
where $\ket{\pm}$ and $\ket{N}$ are the uncoupled eigenstates of $\sigma_z$ and the field respectively.    The corresponding eigenenergies are $E^{\pm}_N=N\hbar\omega_{\mu} - g^2/\hbar\omega_{\mu} \mp \frac{1}{2}E_{Ch}$.  If we now allow $E_J$ to be finite, charge states are mixed and $H$ is no longer diagonal.  In the limit $N \approx \langle N \rangle \equiv \langle a^{\dagger}a \rangle \gg 1$, the matrix elements of the Josephson term, which are all off-diagonal, are 
\begin{displaymath}
\bra{\pm;N-m}(- \frac{1}{2}E_J\sigma_x) \ket{\mp;N} = - \frac{1}{2}E_J J_{\mp m}(\alpha) 
\end{displaymath}
where $J_m(\alpha)$ is the $m$th order Bessel function of the first kind and $\alpha = 4g\sqrt{\langle N \rangle}/\hbar\omega_{\mu}=2E_Q n_{\mu}/\hbar\omega_{\mu}$ is the dimensionless microwave amplitude.   The matrix representation of this dressed Hamiltonian is exact (except that the off-diagonal elements deviate from Bessel functions for small $\langle N \rangle$ \cite{DisplacedNumber}) and is equivalent to the semiclassical Floquet Hamiltonian in the nonuniformly rotating frame of \cite{PeggSeries} (also used by \cite{MITLZ}).  When the multiphoton resonance condition $m\hbar\omega_{\mu} \approx E_{Ch}$ is satisfied for some photon number $m$, two of our dressed states are nearly degenerate.  The first order solution for these levels can then be found by solving the reduced $2\times2$ matrix spanning the degenerate subspace \cite{Salwen}.  The energy bands
\begin{equation}
\label{bands}
E_{\pm m} = \pm \sqrt{\left(E_{Ch}-m\hbar\omega_{\mu}\right)^2 + \left(\frac{E_J}{2} J_{m}(\alpha)\right)^2}
\end{equation}
describe an avoided level crossing with dressed gap $\Delta_m(\alpha) = E_J J_{ m}(\alpha)$ (see Fig. \ref{ManyEj}a).  To calculate the response of the dressed system, we average over different values of $N$ assuming the driving field is in a coherent state characterized by $\langle N \rangle$. 

This energy diagram, shown in Fig. \ref{ManyEj} a, is significantly different from what is found for transverse driving \cite{CohenTann}.  If the transverse field is circularly polarized, only 1-photon transitions ($m=1$) are allowed at all orders of perturbation theory.  For a linearly polarized drive, multiphoton transitions are allowed at higher orders, but only for odd values of $m$.  Even crossings remain degenerate.  For the LDS, the perturbing $\sigma_x$ term allows transitions for all photon numbers at \textit{first order}.

We measure the dressed states by coupling the driven SCB to a rf oscillator.  We probe the oscillator with a weak rf field and measure the magnitude and phase of the reflected signal.  For the measurements, we simultaneously apply the microwave field and rf probe, while slowly sweeping $n_g$ (197 Hz).  The sample is mounted on the mixing chamber of our dilution refrigerator, with a base temperature below 20 mK.  All measurement lines are heavily filtered and attenuated, with the last stage of filtering/attenuation on the mixing chamber.  The microwave power at the sample, after 60 dB of cold attenuation, ranges from nominally \mbox{-80} dBm to \mbox{-50} dBm.  The rf power is nominally -132 dBm, corresponding to roughly $n_{rf} \sim 0.01$ with an average of $N_{osc}\sim6$ rf photons in the oscillator. The rf signal is coupled through a cryogenic circulator mounted on the dilution unit, and the reflected signal is amplified by a cryogenic amplifier at 4 K.  At room temperature, the signal is further amplified and fed into an Aeroflex 3030A vector rf digitizer.  

We find that we can fully explain the rf response by formally treating the dressed states as coherent, tunable two-level systems described by (\ref{bands}).  In Fig. \ref{MagPhase}a, we clearly observe absorption of the rf probe when the dressed states are resonant. We can use this basic phenomenon to quantitatively map the energy diagram of the dressed states, as is illustrated in Fig. \ref{ManyEj}.  We see that the agreement between the data and the theory is quite good.

We can also interpret the phase response, shown in Fig. \ref{MagPhase}b, in terms of the dispersive coupling between the rf oscillator and our tunable two-level systems.  In particular, we see that when the 2nd lobe of each $\Delta_m$ is above the probe frequency plane, there is a clear phase shift, similar to the quantum capacitance response visible in the 0-photon line at low $A_{\mu}$ \cite{SillanpaaQC, DutyQC, ChalmQCTheory}.  This indicates the dispersive shift of the oscillator's frequency caused by the dressed states.  The phase shifts in the 1st lobes have more structure, even changing sign in some regions.  We believe this can be explained by considering the various relaxation processes in the dressed SCB.  For smaller values of $A_{\mu}$, relaxation of the undressed charge states dominates, which mixes the populations of the dressed states, even leading to population inversion \cite{CohenTann}.  For higher $A_{\mu}$, the relaxation also becomes dressed, and the system stays in the ground dressed state. Finally, in \cite{YaleCavity}, sharp, resonant phase features were observed for $\Delta_m < hf_c$.  These are washed out by charge noise in our case because the size of the features, proportional to $\Delta_m/E_Q$, is much smaller.

\begin{figure}
\includegraphics[width=1\columnwidth]{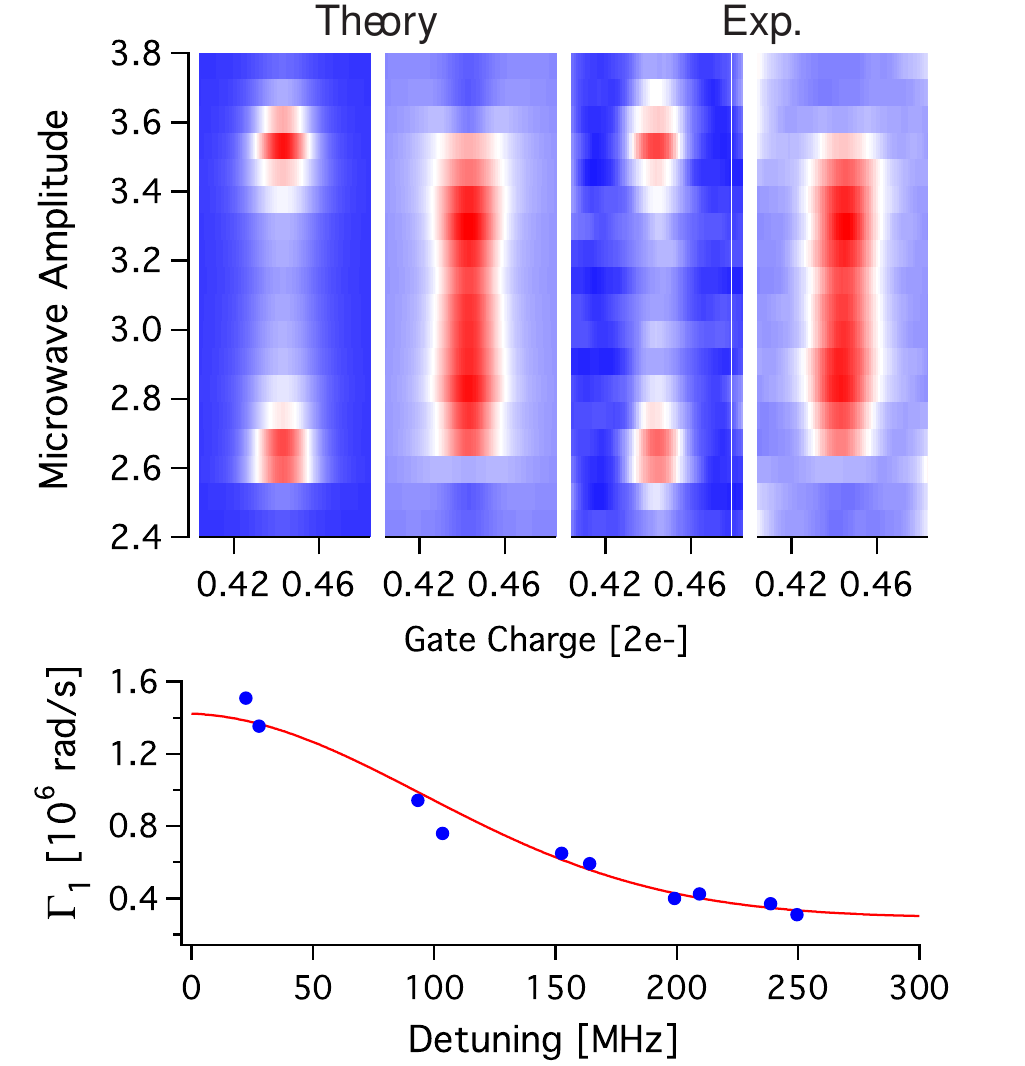}
\caption{(Color Online) Bloch response of the dressed states.  The color scale is the same as Fig. \ref{MagPhase}.  We model the rf response of the the driven SCB by solving the Bloch equations \cite{Bloch} for a tunable two-level system, the dressed state described by (\ref{bands}), driven by the rf probe.  From left to right, we have the theoretical magnitude and phase responses and the corresponding experimental data.  In response to the rf drive voltage, $V_{rf}$, we calculate the in-phase and quadrature component of the dressed state charge, $Q_i$ and $Q_q$ respectively.  We then take $C_{eff} = Q_i/V_{rf}$ and define an effective resistance $R_{eff} =  V_{rf}/\dot{Q_q} = V_{rf}/2\pi f_c Q_q$.  We then compute the rf reflection coefficient, $\Gamma$, including the impedance transformation of the oscillator.  The important fitting parameters are the relaxation rate, $\Gamma_1^{deg}$, and the dephasing rate, $\Gamma_{\varphi}^{deg}$, at degeneracy.  We note that it is not possible to extract unique values for $\Gamma_1^{deg}$ and $\Gamma_{\varphi}^{deg}$ unless the magnitude and phase response are fit simultaneously.  The bottom panel shows the extracted values of $\Gamma_1^{deg}$ as a function of the detuning between the dressed state and oscillator, for positive detuning, giving values of $1/\Gamma_1^{deg} \approx$ 0.7-3.2 $\mu s$.  The fit is explained in the text.  We also extract a value of $1/\Gamma_{\varphi}^{deg} \approx 48$ ns (see text).}
\label{BlochResponse}
\end{figure}

In fact, we can quantitatively explain both the magnitude and phase response by solving the Bloch equations \cite{Bloch} for the dressed state in the vicinity of a dressed degeneracy (see Fig. \ref{BlochResponse}).  We assume that the coherence rates have the standard gate dependence \cite{MakhlinReview} , $\Gamma_1 = \Gamma_1^{deg} \sin^2 \theta$ and $\Gamma_{\varphi} = \Gamma_{\varphi}^{deg}+  \Gamma_{\varphi}^{inf} \cos^2 \theta$, where $\tan\theta = \Delta_m/(E_{Ch}-m\hbar\omega_{mu})$.  (The actual parameter that enters the Bloch equations is $\Gamma_2 = \Gamma_1/2 + \Gamma_{\varphi}$.)  We loosely fit $1/\Gamma_{\varphi}^{inf} = 0.5 \textrm{ ns}$, consistent with previous measurements \cite{ChalmersOsc}, and then let the other parameters vary.  The fits also includes the effect of low-frequency fluctuations of $n_g$, which we model as Gaussian broadening of the Bloch response. We extract a value of $\sigma_n \approx 0.009$ Cooper-pairs, consistent with typical values of 1/f noise in a SCB \cite{NEC1overf}.  Overall, we see that the theory accurately reproduces the rf response, demonstrating the physical nature of the dressed states as two-level systems.

In the bottom pane of Fig. \ref{BlochResponse} we plot the extracted values of $\Gamma_1^{deg}$ as a function of the detuning between the dressed state and the oscillator, $\delta\omega$.   We find $1/\Gamma_1^{deg} \approx$ 0.7-3.2 $\mu s$, with $\Gamma_1^{deg}$ increasing significantly for small detuning.   This is expected since $\Gamma_1^{deg}$ is proportional to the spectral density of the environmental voltage noise, which the oscillator greatly enhances in its band.  The fit represents the standard theory for $\Gamma_1^{deg}$ with an added constant background rate\cite{MakhlinReview}.  The magnitude of the rate agrees well with the theory, extracting a coupling constant $C_{\mu}/C_{\Sigma} \approx 0.02$ which agrees with measured circuit parameters.  We include in the fit the effects of inhomogenous broadening of the dressed state resonance, which can be caused by photon number fluctuations in the oscillator \cite{YaleACstark}.  To estimate this effect, we start with the frequency shift per photon $g_{osc}^2/\pi\delta\omega \approx 26$ MHz, where $g_{osc}/2\pi \approx 15$ MHz is the qubit-oscillator coupling and we have used $\delta\omega = 2\pi f_c/Q$.  Assuming poissonian fluctuations with $\sqrt{N_{osc}} \approx 2.4$, we get a frequency width of $\sim 64$ MHz, which is consistent with the value of $\sim 90$ MHz extracted from the fit.  There can also be a significant contribution to the broadening due to charge noise.   These values are also consistent with the value of $1/\Gamma_{\varphi}^{deg} \approx 48$ ns extracted from the Bloch response.  There are possibly large systematic errors in these values, since the value of  $\Gamma_1^{deg}$ depends sensitively on the rf probe amplitude, which is poorly determined.  If a larger value of the amplitude is assumed, we extract a larger value for $\Gamma_1^{deg}$ and a smaller value for $\Gamma_{\varphi}^{deg}$.  Still, the dependence of $\Gamma_1^{deg}$ on detuning is robust.  These values for $\Gamma_1^{deg}$ and $\Gamma_{\varphi}^{deg}$ are similar to values measured for undressed qubits, especially when the expected scaling with $\Delta_m$ is considered.  This is promising for the prospect of using dressed states as tunable qubits.

In conclusion, we have created LDS of a nanofabricated artificial atom over a wide range of drive strengths.  We have directly measured the physical properties of these states, finding an agreement with theory which is remarkable for a solid-state system.

We would like to thank Enrique Solano, and the members of the Quantum Device Physics and Applied Quantum Physics groups for useful discussions.  The samples were made at the nanofabrication laboratory at Chalmers. The work was supported by the Swedish VR and SSF,  the Wallenberg foundation, and by the European Union under the integrated project EuroSQIP.

\end{document}